\documentclass{llncs}

\usepackage{amssymb}
\usepackage{mdwlist}
\usepackage{graphicx}
\usepackage{amsmath}
\usepackage{threeparttable}
\usepackage{booktabs}
\usepackage{enumitem}
\usepackage{url}
\newcommand{\keywords}[1]{\par\addvspace\baselineskip
\noindent\keywordname\enspace\ignorespaces#1}

\begin{document}

\mainmatter  

\title{On the Security of an enhanced short signature scheme}

\titlerunning{}

%
%
\author{Miaomiao Tian\inst{1,2}%
\thanks{Corresponding author. E-mail: miaotian@mail.ustc.edu.cn (M. Tian).}%
\and Liusheng Huang\inst{1,2} \and Wei Yang\inst{1,2}}

\authorrunning{}

\institute{School of Computer Science and Technology, University of Science and Technology of China, Hefei 230026, P. R. China\\
\and Suzhou Institute for Advanced Study, University of Science and Technology of China, Suzhou 215123, P. R. China\\
}

%
%

\toctitle{}
\tocauthor{}
\maketitle

\begin{abstract}
Currently, short signature is receiving significant attention since it is particularly useful in low-bandwidth communication environments. However, most of the short signature schemes are only based on one intractable assumption. Recently, Su presented an identity-based short signature scheme based on knapsack and bilinear pairing. He claimed that the signature scheme is secure in the random oracle model. Unfortunately, in this paper, we show that his scheme is insecure. Concretely, an adversary can forge a valid signature on any message with respect to any identity in Su's scheme.
\end{abstract}
\section{Introduction}
In traditional public key cryptosystems, a digital certificate which guarantees the authenticity of the relationship between a public key and its owner needs to be produced by a Certification Authority (CA). It brings the certificate management problem since such a system requires a large amount of computing and storage cost to deal with distribution, verification, renewal and storage of the certificates. To overcome this problem, Shamir \cite{Sha84} introduced the concept of identity-based (ID-based) public key cryptography in 1984. In this setting, a user's public key can be determined by his identity (e.g., his name or email address) and his secret key is generated by a trusted third party called the Private Key Generator (PKG). Therefore, ID-based cryptosystem is more convenient than traditional one and carries great weight in cryptography research community.

Currently, ID-based short signature is receiving significant attention since it is particularly useful in low-bandwidth communication environments. However, most of the existing short signature schemes are only based on one intractable assumption, such as bilinear pairing \cite{BLS01,BB04,ZSS04} and RSA \cite{HW09}. Recently, to enhance the security of short signatures, Su \cite{S11} presented an ID-based short signature scheme based on knapsack and bilinear pairing. He claimed that his scheme is provably secure in the random oracle model. Unfortunately, in this paper, we will show that an adversary can forge a signature on \emph{any} message in Su's signature scheme. That is Su's scheme is not secure.

The rest of this paper is organized as follows. In Section 2, we present some preliminaries used throughout the
paper. We review Su's enhanced short signature scheme in Section 3 and analyze its security in Section 4, respectively. Finally, we conclude this paper in Section 5.
\section{Preliminaries}
\subsection{Bilinear pairing}
Let $\mathbb{G}_1$ and $\mathbb{G}_2$ be two cyclic groups of the same prime order $q$. We will view $\mathbb{G}_1$ as an additive group and $\mathbb{G}_2$ as a multiplicative group. A bilinear pairing is a map $e:\mathbb{G}_1\times\mathbb{G}_1\rightarrow\mathbb{G}_2$ with the following three properties.
\begin{enumerate}
\item Bilinearity: For all $a,b\in \mathbb{Z}$ and $P,Q\in\mathbb{G}_1$, the map $e:\mathbb{G}_1\times\mathbb{G}_1\rightarrow\mathbb{G}_2$ satisfies $e(aP,bQ)=e(P,Q)^{ab}$.
\item Non-degeneracy: There are $P,Q\in\mathbb{G}_1$ such that $e(P,Q)\neq1$.
\item Computability: There exists an efficient algorithm to compute $e(P,Q)$ for all $P,Q\in\mathbb{G}_1$.
\end{enumerate}
\subsection{Knapsack problem}
The knapsack problem or subset-sum problem is to determine, given a set of positive integers $W=\{w_1,w_2,\cdots,w_n,t\}$, whether there is subset of the set $\{w_i\}_{i=1}^n$ such that it sums to $t$. Formally, the problem is equivalent to determining whether there is a set $X=\{x_1,x_2,\cdots,x_n\}$ such that $$\sum_{i=1}^nx_iw_i=t, \quad x_i\in\{0,1\}.$$
\subsection{ID-based signatures}
An ID-based signature scheme consists of the following four probabilistic polynomial-time algorithms:
\begin{description}
\item[\textbf{Setup.}] On input a security parameter $k$, the PKG generates a master secret key $MSK$ and the public parameters $PP$.
\item[\textbf{Extract.}] On input a user's identity $ID$ and the master secret key $MSK$, this algorithm outputs a secret key $sk_{ID}$ for $ID$. The user's public key is determined by his identity $ID$.
\item[\textbf{Sign.}] On input a message $m$ and the secret key $sk_{ID}$ of the signer $ID$, this algorithm outputs a signature $\sigma$ on $m$.
\item[\textbf{Verify.}] On input a signature $\sigma$, a message $m$ and an identity $ID$, it returns 1 if $\sigma$ is a valid signature, and returns 0 otherwise.
\end{description}
\section{Review of Su's signature scheme}
In this section, we review Su's signature scheme \cite{S11}. The scheme is described as follows:
\begin{description}
\item[\textbf{Setup.}] Given a security parameter $k$, the PKG chooses two groups $\mathbb{G}_1$ and $\mathbb{G}_2$ of the same prime order $q$ as well as a bilinear map $e:\mathbb{G}_1\times\mathbb{G}_1\rightarrow\mathbb{G}_2$. It also chooses a random generator $P$ of $\mathbb{G}_1$, the master secret key $s\in \mathbb{Z}^*_q$ and a hash function $H:\{0,1\}^*\rightarrow\mathbb{Z}^*_q$. Afterwards, the PKG sets $Q_S=sP$ as the master public key of the system and publishes the public parameters $PP=(\mathbb{G}_1,\mathbb{G}_2,e,P,Q_S,H)$.
\item[\textbf{Extract.}] On input the master secret key $s$ and an identity $ID\in\{0,1\}^*$, the user with identity $ID\in\mathbb{Z}^*_q$ chooses a random value $sk_{ID}\in\mathbb{Z}^*_q$ as his secret key and publishes his public key $Q_{ID}=ID\times sk_{ID}\times Q_S$.
\item[\textbf{Sign.}] On input a message $m\in\{0,1\}^*$, the signer $ID$ with private key $sk_{ID}$ does the following steps:
\begin{enumerate*}
\item Choose two random vectors $B=(b_1P,b_2P,\cdots,b_nP)$ and $X=(x_1,x_2,\cdots,x_n)$, where $b_i\in\mathbb{Z}^*_q$, $x_i\in\{0,1\}$.
\item Compute $a_i=b_i\times sk_{ID} \pmod q$ for each $i\in\{1,\cdots,n\}$.
\item Compute $U=\sum_{i=1}^{n}x_ib_iP$ and $V=\lambda\sum_{i=1}^{n}x_ia_iP$, where $\lambda=H(m)$.
\item Output the signature $\sigma=(U,V)$.
\end{enumerate*}
\item[\textbf{Verify.}] On input a signature $\sigma=(U,V)$, a message $m$ and an identity $ID$ as well as the corresponding public key $Q_{ID}$, a verifier computes $\lambda=H(m)$ and then checks if $$e(U,Q_{ID})^\lambda=e(V,Q_S)^{ID}.$$ If so, he accepts the signature; otherwise, he rejects it.
\end{description}

The correctness of the signature can be verified as follows:
\begin{eqnarray*}
         e(U,Q_{ID})^\lambda&=&e(\sum_{i=1}^{n}x_ib_iP,ID\times sk_{ID}\times Q_S)^\lambda  \\
                            &=&e(\sum_{i=1}^{n}x_ia_iP,ID\times Q_S)^\lambda  \\
                            &=&e(\lambda\sum_{i=1}^{n}x_ia_iP,Q_S)^{ID}  \\
                            &=&e(V,Q_S)^{ID}
\end{eqnarray*}
\textbf{Remark 1.} We can see that, in Su's signature scheme, a verifier cannot confirm a user's public key from its identity since the user's public key involves a random secret value. Therefore, Su's signature scheme is not a standard ID-based signature scheme.\\
\textbf{Remark 2.} We also note that Su's scheme is not a certificateless signature scheme \cite{AP03}. In a certificateless signature scheme, the secret key of a user is a combination of his partial private key generated by the PKG and some secret value chosen by the user himself.
\section{Security analysis on Su's signature scheme}
Su \cite{S11} claimed that his signature scheme is secure in the random oracle model. However, he didn't give a formal proof of the scheme. In this section, we will present two forgery attacks on his signature scheme and show that his scheme is insecure. In the first attack, a polynomial time adversary $\mathcal{A}$, who has received a valid signature with respect to the user with identity $ID$, can forge a signature on \emph{any} new message for the same user $ID$. In the second attack, we will show that the adversary $\mathcal{A}$ is also able to forge a signature on \emph{any} message for a new user $ID'$.
\subsection{Attack I}
Assume that the adversary $\mathcal{A}$ aims to forge a signature on $m^*$ with respect to the user with identity $ID$. After receiving $ID$'s valid signature $\sigma=(U,V)$ on message $m$. $\mathcal{A}$ makes two hash queries to obtain the values $\lambda=H(m)$ and $\lambda^*=H(m^*)$. Then he sets $U^*=U$ and $V^*=\frac{\lambda^*}{\lambda}V$, and outputs $\sigma^*=(U^*,V^*)$ as a signature on $m^*$. We can verify its validity as follows:
\begin{eqnarray*}
                        e(V^*,Q_S)^{ID}&=&e(\frac{\lambda^*}{\lambda}V,Q_S)^{ID}  \\
                            &=&\big(e(V,Q_S)^{ID}\big)^{\frac{\lambda^*}{\lambda}}  \\
                            &=&\big(e(U,Q_{ID})^\lambda\big)^{\frac{\lambda^*}{\lambda}}  \\
                            &=&e(U,Q_{ID})^{\lambda^*}   \\
                            &=&e(U^*,Q_{ID})^{\lambda^*}
\end{eqnarray*}
\subsection{Attack II}
Due to there is no authentication information for a user's public key, an adversary can replace any user's public key with a value of his choice. Here we show that the adversary $\mathcal{A}$ is also able to forge a signature on $m^*$ with respect to another identity $ID'$.

After receiving $ID$'s valid signature $\sigma=(U,V)$ on message $m$. $\mathcal{A}$ makes two hash queries to obtain the values $\lambda=H(m)$ and $\lambda^*=H(m^*)$. Then he sets $Q_{ID'}=\frac{ID'}{ID}Q_{ID}$, $U^*=U$ and $V^*=\frac{\lambda^*}{\lambda}V$, and outputs $\sigma^*=(U^*,V^*)$ as a signature on $m^*$ with respect to the user $ID'$. We can verify the validity of the signature as follows:
\begin{eqnarray*}
                        e(V^*,Q_S)^{ID'}&=&e(\frac{\lambda^*}{\lambda}V,Q_S)^{ID'}  \\
                            &=&\big(e(V,Q_S)^{ID}\big)^{\frac{ID'\times\lambda^*}{ID\times\lambda}}  \\
                            &=&\big(e(U,Q_{ID})^\lambda\big)^{\frac{ID'\times\lambda^*}{ID\times\lambda}}  \\
                            &=&e(U,\frac{ID'}{ID}Q_{ID})^{\lambda^*}   \\
                            &=&e(U^*,Q_{ID'})^{\lambda^*}
\end{eqnarray*}

The two attacks demonstrate that, in Su's signature scheme, anyone can forge a valid signature on \emph{any} message with respect to \emph{any} identity as long as he has obtained one valid signature. Therefore, Su's signature scheme is not secure.
\section{Conclusion}
Recently, Su \cite{S11} presented an enhanced short signature scheme and claimed that it is secure in the random oracle model. In this paper, however, we have demonstrated that an adversary can forge a signature on \emph{any} message with respect to \emph{any} identity. In other words, Su's signature scheme is not secure.
\section*{Acknowledgements}
This work is supported by the Major Research Plan of the National Natural Science Foundation of China (No. 90818005), the National Natural Science Foundation of China (No. 60903217), the Fundamental Research Funds for the Central Universities (No. WK0110000027), and the Natural Science Foundation of Jiangsu Province of China (No. BK2011357).









\begin{thebibliography}{00}
\bibitem{Sha84} Shamir A. Identity-based cryptosystems and signature schemes. In: Proceedings of advances in cryptology--CRYPTO'85. LNCS, vol. 196. Springer-Verlag; 1985. p. 47--53.
\bibitem{BLS01} Boneh D, Lynn B, Shacham H. Short Signatures from the Weil Pairing. In: Proceedings of advances in cryptology--ASIACRYPT 2001. LNCS, vol. 2248. Springer-Verlag; 2001. p. 514--532.
\bibitem{BB04} Boneh D, Boyen X. Short Signatures Without Random Oracles. In: Proceedings of advances in cryptology--EUROCRYPT 2004. LNCS, vol. 3027. Springer-Verlag; 2004. p. 56--73.
\bibitem{ZSS04} Zhang F, Safavi-Naini R, Susilo W. An Efficient Signature Scheme from Bilinear Pairings and Its Applications. In: Proceedings of the 7th International Workshop on Theory and Practice in Public Key Cryptography (PKC'04). LNCS, vol. 2947. Springer-Verlag; 2004. p. 277--290.
\bibitem{HW09} Hohenberger S, Waters B. Short and stateless signatures from the RSA assumption. In: Proceedings of advances in cryptology--CRYPTO 2009. LNCS, vol. 5677. Springer-Verlag; 2009. p. 654--670.
\bibitem{S11} Su PC. Enhanced short signature scheme with hybrid problems. Computers \& Electrical Engineering 2011;37(2):174--179.
\bibitem{AP03} Al-Riyami S, Paterson K. Certificateless public key cryptography. In: Proceedings of advances in cryptology--ASIACRYPT 2003. LNCS, vol. 2894. Springer-Verlag; 2003. p. 452--473.

\end{thebibliography}
\end{document}